\documentclass[english,journal=jpcafh,manuscript=article,email=true,etalmode=truncate,maxauthors=0,doi=true]{achemso}

\SectionNumbersOn

\usepackage[version=3]{mhchem} 
\usepackage{braket} 

\usepackage{hyperref} 
\hypersetup{colorlinks=true, citecolor=blue, urlcolor=blue, linkcolor=blue}
\usepackage{cleveref}
  	\crefname{figure}{Figure}{Figures}
  	\crefname{table}{Table}{Tables}
  	\crefname{equation}{Eq.}{Eqs.}
  	\crefname{section}{Section}{Sections}
  	\crefname{subsection}{Section}{Sections}
  	\crefname{subsubsection}{Section}{Sections}
  	\crefname{algorithm}{Algorithm}{Algorithms}

\newcommand{\doi}[1]{\href{http://dx.doi.org/#1}{\nolinkurl{#1}}}

\usepackage{todonotes}

\newcommand*{\bigo}[1]{\ensuremath \mathcal{O}(#1)}
\newcommand{\code}[1]{\texttt{#1}}

\author{Nakul K. Teke}
\author{Ajay Melekamburath}
\author{Bimal Gaudel}
\author{Edward F. Valeev}
\email{efv@vt.edu}
\affiliation[Virginia Tech]
{Department of Chemistry, Virginia Tech, Blacksburg, VA 24061}
\title[Best way to treat triples in coupled cluster?]
  {``Best'' iterative coupled-cluster triples model: More evidence for 3CC.}

\keywords{coupled cluster, CCSDT, 3CC, thermochemistry}

\begin{document}







\begin{abstract}
To follow up on the unexpectedly-good performance of several coupled-cluster models with approximate inclusion of 3-body clusters [{\em J. Chem. Phys.} {\bf 151}, 064102 (2019)] we performed a more complete assessment of the 3CC method [{\em J. Chem. Phys.} \textbf{125}, 204105 (2006)] for accurate computational thermochemistry in the standard HEAT framework. New spin-integrated implementation of the 3CC method applicable to closed- and open-shell systems utilizes a new automated toolchain for derivation, optimization, and evaluation of operator algebra in many-body electronic structure. We found that with a double-zeta basis set the 3CC correlation energies and their atomization energy contributions are almost always more accurate (with respect to the CCSDTQ reference) than the CCSDT model as well as the standard CCSD(T) model. The mean absolute errors in cc-pVDZ \{ 3CC, CCSDT, and CCSD(T) \} electronic (per valence electron) and atomization energies relative to the CCSDTQ reference for the HEAT dataset [{\em J. Chem. Phys.} {\bf 121}, 11599 (2004)], were \{24, 70, 122\} $\mu E_\mathrm{h}/e$
and \{0.46, 2.00, 2.58\} kJ/mol,
respectively. The mean absolute errors in the complete-basis-set limit \{ 3CC, CCSDT, and CCSD(T) \} atomization energies relative to the HEAT model reference, were \{0.52, 2.00, and 1.07\} kJ/mol, The significant and systematic reduction of the error by the 3CC method and its lower cost than CCSDT suggests it as a viable candidate for post-CCSD(T) thermochemistry applications, as well as the preferred alternative to CCSDT in general.
\end{abstract}

\section{Introduction}
The coupled-cluster model\cite{VRG:coester:1958:NP, VRG:coester:1960:NP, VRG:cizek:1966:JCP, Paldus_1972} is the gold standard of molecular electronic structure theory, with a growing footprint in materials\cite{AM:Rosciszewski:2000:PRB, AM:Voloshina:2011:PCCP, AM:Booth:2013:N, AM:Sansone:2016:JPCL, AM:McClain:2017:JCTC, AM:Zhang:2019:FM} and nuclear\cite{AM:Dean:2004:PRC, AM:Baardsen:2013:PRC, AM:Hagen:2014:RPP, AM:Sahoo:2020:NJP, AM:Novario:2021:PRL} physics. It allows compact and systematically improvable description of electron correlation in states dominated by a single determinant and serves as a versatile platform for description of multiconfiguration electronic states within the reference and the adjacent sectors of the Fock space.\cite{VRG:krylov:2008:ARPC,VRG:musial:2008:JCP,VRG:lyakh:2011:CR} Inclusion of higher-body correlators allows for small molecules to rival the uncertainties in experimentally-derived chemical energetics,\cite{VRG:tajti:2004:JCP, VRG:boese:2004:JCP, VRG:bomble:2006:JCP, VRG:harding:2008:JCP, VRG:sylvetsky:2016:JCP}
with reduced-scaling\cite{VRG:riplinger:2013:JCP, VRG:pavosevic:2014:JCP, VRG:riplinger:2016:JCP, VRG:pavosevic:2017:JCP, VRG:peng:2018:JCTC, AM:Crawford:2019:WCMS} and adaptive\cite{AM:Lyakh:2010:JCP,VRG:xu:2018:PRL} CC extensions promising similar accuracy for larger systems.

In practice applications of CC methods utilize {\em heuristic} CC-based models that balance cost and accuracy. A key example of such a heuristic is the traditional CCSD(T) model\cite{VRG:raghavachari:1989:CPL,VRG:watts:1993:JCP,VRG:stanton:1997:CPL} that combines self-consistent (iterative) treatment of CC 1- and 2-body correlators (singles and doubles, respectively) with perturbative (non-iterative) treatment of 3-body correlators (triples) at $\bigo{N^7}$ cost, with $N$ proportional to system size.
CCSD(T)'s accuracy for chemical energetics {\em significantly} exceeds\cite{AM:Mardirossian:2017:MP} that of even the best empirical Kohn-Sham density functional theory (KS DFT) methods,\cite{VRG:kohn:1965:PR, AM:Parr:1995:ARPC} thus warranting its much higher computational cost than the $\bigo{N^4}$ cost of DFT.
Although CCSD(T) is usually viewed as an economical approximation to the exact CCSDT model whose cost is $\bigo{N^8}$, CCSD(T) is usually more accurate than CCSDT,\cite{VRG:feller:2006:JCP} thus from the practical standpoint to warrant post-CCSD(T) treatment
it would be mandatory to include quadruples, at $\bigo{N^9}$  cost\cite{VRG:bomble:2005:JCP} or higher. This partially explains why CCSD(T) is indeed a ``sweet spot'' in the cost-to-accuracy sense and why the majority of lower-end thermochemistry models stop at (T).\cite{VRG:parthiban:2001:JCP,VRG:wood:2006:JCP,VRG:deyonker:2006:JCP,VRG:curtiss:2007:JCP}  
Efficient reduced-scaling formulations (asymptotically reaching linear scaling) of CCSD(T),\cite{VRG:pinski:2015:JCP,VRG:riplinger:2016:JCP,VRG:ma:2018:JCTC,VRG:nagy:2018:JCTC} in combination with explicit correlation (F12),\cite{VRG:pavosevic:2017:JCP,VRG:ma:2017:JCTC,VRG:kumar:2020:JCP} have extended its useful application range of CCSD(T) to molecules with hundreds to atoms.

Unfortunately CCSD(T), despite often referred to as the ``gold standard'' of quantum chemistry, is not sufficient for practical applications in several scenarios, such as (a) when high accuracy is needed, such as for predictive chemical energetics, or (b) when a single determinant is no longer a good reference.
While other $\bigo{N^7}$ heuristics exist that can improve on CCSD(T),\cite{VRG:crawford:1998:IJQC,VRG:hirata:2004:JCPa,VRG:piecuch:2005:JCP,VRG:taube:2008:JCP} efficient self-consistent treatment of triples {\em and} higher-body correlators is necessary to address the shortcomings of CCSD(T). 
The importance of iterative treatment of triples and higher-body clusters for accurate chemical energetics for reaching even chemical accuracy (1 kcal/mol) has been long recognized in the design of models for high-end thermochemistry (HEAT,\cite{VRG:tajti:2004:JCP,VRG:bomble:2006:JCP,VRG:harding:2008:JCP} W3+\cite{VRG:boese:2004:JCP,VRG:karton:2006:JCP},   FPD\cite{VRG:feller:2006:JCP,VRG:feller:2008:JCP}) protocols.
Clearly, to make high-order CC applications practical it will be necessary to develop adaptive (to control the accuracy) and numerically-efficient (to control the cost) variants thereof. While such developments are a current research frontier for many groups,\cite{VRG:xu:2018:PRL} a relevant shorter-term question whether useful higher-order CC heuristics exist that allow to reach the {\em ab initio} limit more economically.

In the course of designing a higher-order extension of one such heuristic\cite{ VRG:rishi:2019:JCP} (namely, a  distinguishable cluster\cite{VRG:kats:2013:JCPa} approximation to CCSDT) some of us stumbled on one promising heuristic, the so-called 3CC method introduced by Bartlett and Musia{\l}.\cite{VRG:bartlett:2006:JCP}
For a small set of molecules the 3CC method was found to produce correlation energies significantly closer to the reference CCSDTQ values (obtained at the $\mathcal{O}(N^{10})$ cost) than its parent CCSDT method, while its performance for bond-breaking was found to be similar to that of CCSDT. Comparison to other CCSDT approximations (e.g., DCSDT and pCCSDT, both introduced in Ref. \citenum{VRG:rishi:2019:JCP}) was also favorable. Unfortunately, the original study\cite{VRG:rishi:2019:JCP} was rather limited in scope (e.g., only closed-shell systems were considered). The goal of this study is to assess the performance of the 3CC method for a larger set of closed- and open-shell systems in the HEAT benchmark set\cite{VRG:tajti:2004:JCP,VRG:harding:2008:JCP} (for which definitive reference energies are available) with an eye towards making accurate thermochemistry models limited to CC triples more practical.

This article is organized as follows. The  formalism of the CC method and its 3CC heuristic approximation are briefly recapitulated in \cref{sec:formalism}; for a more thorough introduction to the rich phenomenology of the $n$CC models and their cousins the reader is referred to the original literature. \cite{VRG:bartlett:2006:JCP,VRG:musial:2007:JCP}
Technical details of the automated implementation spin-free and spin-integrated closed- and open-shell variants of CCSDT and 3CC methods are described in \cref{sec:technical-details}.
Assessment of the 3CC method using the HEAT thermochemistry benchmark is reported in \cref{sec:results} and our findings are summarized in \cref{sec:summary}.

\section{Formalism}
\label{sec:formalism}

\subsection{The 3CC model}
\label{sec:formalism-3cc}

The coupled-cluster wave function is obtained from the reference determinant $\ket{0}$ by the action of the exponentiated cluster operator, $\hat{T}$,
\begin{align}
  \label{eq:psi}
   \ket{\Psi_{\text{CC}}} \equiv \exp(\hat{T}) \ket{0},
  \end{align}
that conventionally includes up to $K$-body correlators (cluster operators),
\begin{align}
\label{eq:T}
    \hat{T} = \hat{T}^{(K)} \equiv \sum_{k=1}^K \hat{T}_k,
\end{align}
with a $k$-body correlator defined in standard (non-unitary) as
\begin{align}
\label{eq:Tk}
    \hat{T}_k \equiv \frac{1}{(k!)^2}t_{a_1\dots a_k}^{i_1\dots i_k} a^{a_1\dots a_k}_{i_1\dots i_k} .
\end{align}
Here we use the standard tensor notation of many-body quantum chemistry\cite{VRG:harris:1981:PRA,VRG:kutzelnigg:1982:JCP} 
\footnote{A recent comprehensive description of the tensor notation for electronic structure can be found in Ref. \citenum{VRG:calvin:2020:CR}.} whereby tensor elements and operators are written in covariant form, with \{sub,super\}scripts denoting \{bra,ket\} or \{annihilation,creation\} indices, respectively. In \cref{eq:Tk} replacement operator given in tensor form is defined in terms of fermionic creators ($\hat{a}^\dagger$) and annihilators ($\hat{a}$) as $a^{a_1\dots a_k}_{i_1\dots i_k} \equiv a^{\dagger}_{a_1} \dots a^{\dagger}_{a_k} a_{i_k}\dots a_{i_1}$, hence tensor operator $a^{a_1\dots a_k}_{i_1\dots i_k}$ is antisymmetric with respect to arbitrary permutations of bra or ket indices; by convention, the $t$ amplitudes in \cref{eq:Tk} are similarly antisymmetric.
As is traditional, $ijk\dots$ and $abc\dots$ will refer to the spinorbitals present (occupied) and missing (unoccupied) in $\ket{0}$, respectively.
The CC model is defined by our choice of $K$ in \cref{eq:T}: $K=2$ for CCSD,\cite{VRG:purvis:1982:JCP} $K=3$ for CCSDT,\cite{VRG:noga:1987:JCP} $K=4$ for CCSDTQ,\cite{VRG:kucharski:1992:JCP} etc.

In {\em traditional} CC, the amplitudes are determined by projection of the CC Schr\"odinger equation, $\hat{H} \ket{\Psi_{\text{CC}}} = E \ket{\Psi_{\text{CC}}}$, onto the biorthogonal CC excited manifolds, $\bra{0}a_{a_1\dots a_k}^{i_1\dots i_k} \exp(-\hat{T})$, of rank $k\leq K$:
\begin{align}
\label{eq:t-eqs}
\bra{0}a_{a_1\dots a_k}^{i_1\dots i_k} \exp(-\hat{T}) \hat{H} \exp(\hat{T}) \ket{0} \equiv \bra{0}a_{a_1\dots a_k}^{i_1\dots i_k} \bar{H} \ket{0} = 0, \quad \forall k\in[1,K]
\end{align}
where
\begin{align}
\label{eq:hbar}
\bar{H} \equiv& \exp(-\hat{T}) \hat{H} \exp(\hat{T}) \nonumber \\
= & \hat{H} + [\hat{H}, \hat{T}] + \frac{1}{2!} [[\hat{H}, \hat{T}], \hat{T}] + \frac{1}{3!} [[[\hat{H}, \hat{T}], \hat{T}], \hat{T}] + \frac{1}{4!} [[[[\hat{H}, \hat{T}], \hat{T}], \hat{T}], \hat{T}].
\end{align}
The CC energy is obtained as
\begin{align}
\label{eq:cc-energy}
    E = \bra{0} \bar{H} \ket{0}.
\end{align}

The 3CC model which is our focus here is a member of the $n$CC family of models ($n=2,3,\hdots$) introduced by Bartlett and Musia{\l}.\cite{VRG:bartlett:2006:JCP,VRG:musial:2007:JCP}
The $n$CC family, one of several known families of {\em internally-corrected}\cite{AM:Paldus:2017:JMC} coupled-cluster methods, were designed to be exact for an $n$-electron system (or any number of noninteracting instances of such systems)
by removing from the CC amplitude equations (\cref{eq:t-eqs}) the contributions quadratic in 2- and higher-body clusters that have so-called non-hole-conjoined (NHCJ) structure. Unlike their hole-conjoined (HCJ) counterparts, the NHCJ counterparts do not contribute to the cancellation of the energy dependent terms in the corresponding truncated CI amplitude equations for an $n$-electron system. Specifically, for the $n=2$ case the following contributions to the doubles equations are omitted:
\begin{align}
    \label{eq:2cc-delta}
    {\hat{A}^{{a_1}{a_2}}_{{i_1}{i_2}}} \left( {{{\frac{1}{2}}}{\bar{g}^{{a_3}{a_4}}_{{i_3}{i_4}}}{t^{{i_1}{i_3}}_{{a_1}{a_3}}}{t^{{i_2}{i_4}}_{{a_2}{a_4}}}}
    {{{-\frac{1}{4}}}{\bar{g}^{{a_3}{a_4}}_{{i_3}{i_4}}}{t^{{i_3}{i_4}}_{{a_1}{a_3}}}{t^{{i_1}{i_2}}_{{a_2}{a_4}}}}
    \right)
    ,
\end{align}
where $\bar{g}^{{a_3}{a_4}}_{{i_3}{i_4}}\equiv g^{{a_3}{a_4}}_{{i_3}{i_4}} - g^{{a_3}{a_4}}_{{i_4}{i_3}}$ is the antisymmetrized Coulomb integral and ${\hat{A}^{{a_1}{a_2}}_{{i_1}{i_2}}}$ is the antisymmetrizer defined as
\begin{align}
{\hat{A}^{{a_1}{a_2}}_{{i_1}{i_2}}} f(i_1,i_2,a_1,a_2) \equiv f(i_1,i_2,a_1,a_2) - f(i_2,i_1,a_1,a_2) - f(i_1,i_2,a_2,a_1) + f(i_2,i_1,a_2,a_1).
\end{align}
There is unfortunately no standard nomenclature for these CC contributions: the two terms in \cref{eq:2cc-delta} correspond (respectively) to the diagrams 1(d) and 1(c) in Ref. \citenum{VRG:bartlett:2006:JCP}, diagrams 1+2 and 3 in Ref. \citenum{VRG:jankowski:1980:IJQC}, diagrams D and C in Refs. \citenum{VRG:huntington:2010:JCP,VRG:rishi:2019:JCP},
and diagrams A+A' and C in Ref. \citenum{VRG:kats:2013:JCP}.

The motivation for exclusion of the NHCJ terms in $n$CC was to find the most compact (hence, economical) and most CI-like method exact for an $n$-electron system but preserving the favorable qualities of truncated CC, namely extensivity and invariance with respect to unitary rotation of occupied/unoccupied orbitals in the traditional form. Note that much earlier Paldus and co-workers motivated the removal of terms in \cref{eq:2cc-delta} from the doubles equations of CCD by the improved description of strongly-correlated systems; the resulting ACP-D45 method\cite{VRG:jankowski:1980:IJQC} is also known as ACCD method of Dykstra and co-workers;\cite{AM:Chiles:1981:CPL} extensions of this idea to CCSD produces the 2CC method and was explored under names ACP(CCSD)\cite{VRG:kucharski:1991:TCAa} and ACCSD.\cite{VRG:piecuch:1996:PRA} For the sake of simplicity we refer to all of these namesakes by its latest name (2CC) to maintain consistent terminology.

Note, however, that 2CC is not the most compact approximation to CCSD that is exact for two or more non-interacting 2-e systems because there are other partial cancellations between diagrams that can be exploited. For example, two  quadratic terms in the doubles equations not involved in \cref{eq:2cc-delta} can be linearly combined to eliminate one of them, as pointed out by Huntington and Nooijen\cite{VRG:huntington:2010:JCP} in the design of their parametrized CCSD method, pCCSD($\alpha$,$\beta$); namely method pCCSD(0,0) contains one fewer diagram than 2CC and seems to perform better for molecular geometries (see Table I of Ref. \citenum{VRG:huntington:2010:JCP}) although neither was deemed to be the most promising variant.\cite{VRG:huntington:2012:JCP}

Note that like its CCSD parent 2CC is exact for a 2-electron system, but unlike CCSD 2CC is not exact for a 2-hole system (e.g. hydrogen fluoride in minimal basis); the breaking of this symmetry also holds for the higher-order $n$CC variants. While this ``symmetry'' was considered in the design of some internally-corrected CC methods\cite{VRG:kats:2013:JCP}, its importance is not obvious, especially for the purposes of computing accurate electron correlation energies since the exact numerical treatment of CC necessarily requires $V \gg O $, with $O/V$ the number of occupied/unoccupied orbitals in the reference.

The key to the design of the 2CC method is that it can be systematically applied to CCSDT and higher-rank methods, yielding 3CC, 4CC, and higher-rank models.
Specifically, the 3CC model obtained by eliminating the following terms from the triples amplitude equation of CCSDT:
\begin{align}
\label{eq:3cc-delta}
\hat{A}_{i_1 i_2 i_3}^{a_1 a_2 a_3} \left( {{{\frac{1}{8}}}{\bar{g}^{{a_4}{a_5}}_{{i_4}{i_5}}}{t^{{i_1}{i_2}}_{{a_1}{a_4}}}{t^{{i_3}{i_4}{i_5}}_{{a_2}{a_3}{a_5}}}} + 
    {{{\frac{1}{24}}}{\bar{g}^{{a_4}{a_5}}_{{i_4}{i_5}}}{t^{{i_4}{i_5}}_{{a_1}{a_4}}}{t^{{i_1}{i_2}{i_3}}_{{a_2}{a_3}{a_5}}}} +
    {{{\frac{1}{4}}}{\bar{g}^{{a_4}{a_5}}_{{i_4}{i_5}}}{t^{{i_1}{i_4}}_{{a_1}{a_4}}}{t^{{i_2}{i_3}{i_5}}_{{a_2}{a_3}{a_5}}}} + 
    {{{\frac{1}{2}}}{\bar{g}^{{a_4}{a_5}}_{{i_4}{a_1}}}{t^{{i_1}{i_2}}_{{a_2}{a_4}}}{t^{{i_3}{i_4}}_{{a_3}{a_5}}}}\right) .
\end{align}
While 3CC is again not the most compact extensive approximation to CCSDT that is still exact for a 3-electron system (further diagrammatic eliminations are possible by using our pCCSDT(0,0,0) method\cite{VRG:rishi:2019:JCP}),
its performance in practice has been found to match or beat that of CCSDT.
Bartlett and Musia{\l} observed it to perform similarly to or better than the full CCSDT for bond breaking benchmarks\cite{VRG:bartlett:2006:JCP} and for excited, ionized, and electron-attached states\cite{VRG:musial:2007:JCP}. We noted its systematic (and significant) improvement over CCSDT and other internally-corrected iterative triples CC methods for correlation energies of several small closed-shell molecules.\cite{VRG:rishi:2019:JCP}

Although formally both CCSDT and 3CC have the same complexity, namely $\bigo{N^8}$, and for medium and large basis sets both are dominated by the $\bigo{O^3V^5}$ contraction involved in the evaluation of the particle-particle ladder (PPL) contribution, there is an important difference in that the latter does not require evaluation of any non-ladder $\bigo{O^4V^4}$ contractions. Therefore even by pure factorization of Hamiltonian it should be possible to make 3CC significantly less expensive than CCSDT. An efficient implementation of 3CC will be considered in \cref{sec:summary} after first examining its performance.

\section{Technical details}\label{sec:technical-details}

In this work we attempt to assess the performance of 3CC for computational thermochemistry more fully. To the best of our knowledge there is no publicly available implementation of 3CC method, and no open-shell reference implementation thereof has been reported. Thus the first task was to implement 3CC model for closed-shell (RHF) and open-shell (UHF) reference states and with adequate efficiency for initial testing on systems with up to 10 atoms. Such development was made possible by several innovations developed in our group:
\begin{itemize}
\item new \code{SeQuant} engine for symbolic tensor algebra that allows to apply Wick's theorem, symbolically simplify (e.g., perform spin integrations), factorize, and interpret (numerically evaluate) the resulting tensor algebra efficiently using an external numerical tensor algebra package,
\item \code{TiledArray} parallel tensor algebra framework\cite{VRG:calvin:2015:I15WIAAA} that supports efficient representation and manipulations of block-sparse distributed tensors, and
\item \code{MPQC} electronic structure package\cite{VRG:peng:2020:JCP} that implements the requisite machinery of electronic structure needed to implement many-body methods.
\end{itemize}
These components form a new automated ``toolchain'' for rapid development of production-quality implementations of many-body methods. While \code{TiledArray} and \code{MPQC} have been described in some detail elsewhere,\cite{VRG:calvin:2015:I15WIAAA,VRG:peng:2020:JCP} the \code{SeQuant} engine has not yet been described (although its source code, including the development branches, is public, just like \code{TiledArray}). The key code innovations of \code{SeQuant} and its integration with \code{TiledArray} that made this work possible will be described elsewhere soon. Meanwhile here we only briefly recap the toolchain's essential features.

\begin{itemize}
\item {\bf Symbolic Operator Algebra.} \code{SeQuant} can be used for efficient symbolic manipulation of algebraic expressions involving tensors over scalar (CC amplitudes, Hamiltonian tensors) and operator (normal-ordered second-quantized operators) fields. Permutational symmetries of tensors and the resulting topological structure of the tensor networks are utilized in the course of symbolic manipulation. Efficient application of Wick's theorem brings products of tensor operators the their canonical form given by a sum of tensor networks, and thus allows evaluation of vacuum expectation values in \cref{eq:t-eqs,eq:cc-energy}. Wick's theorem in general produces expressions that contain redundant (equivalent) terms, thus it is mandatory to combine them before efficient evalutation. An efficient colored-graph-based tensor network canonicalizer in \code{SeQuant} ensures that tensor expressions like \cref{eq:2cc-delta,eq:3cc-delta}  are reduced to their optimal form.
\item {\bf Spin Integration.} Tensor expressions in spin-orbital basis (e.g., \cref{eq:2cc-delta,eq:3cc-delta}) can be used directly, but it is more efficient to perform spin integration symbolically. Symbolic handling of spin also eliminates the need to incorporate the additional logic related to spin quantum numbers in the tensor backend (as done for example in TCE\cite{VRG:hirata::TCA}, \code{libtensor}\cite{VRG:epifanovsky:2013:JCC}, and \code{TAMM}\cite{VRG:mutlu:2023:JCP}). Thus in the course of this work we added the necessary symbolic manipulation capabilities to \code{SeQuant} (contained in \code{SeQuant/domain/mbpt/spin.hpp}). The equations were evaluated in their spin-integrated form for both closed and open-shell molecules, but for efficiency the closed-shell equations were transformed into their biorthogonal form.\cite{VRG:pulay:1984:JCP,VRG:matthews:2013:JCTC,VRG:wang:2018:} Further details of spin integration will be presented in the forthcoming \code{SeQuant} manuscript.
\item {\bf Model-Specific Manipulations.} The truncated traditional CC models and their many approximations (e.g., CCSDT\{1,2,3,4\}, CC3, among others), can be specified fully at the operator level (\cref{eq:t-eqs,eq:hbar,eq:cc-energy}). $n$CC models cannot be specified at the operator level; instead, they are specified by omission of specific tensor networks (terms, diagrams). In principle the specific terms that are omitted in $n$CC methods (\cref{eq:2cc-delta,eq:3cc-delta}) can be identified programmatically, however \code{SeQuant} lacks {\em generic} pattern matching for tensor networks, thus instead such methods are implemented by literally {\em subtracting} the corresponding terms (encoded in C++ source as LaTeX forms of \cref{eq:2cc-delta,eq:3cc-delta} and parsed by \code{SeQuant}) from the CC residual equations in their symbolic form. The analogy with the ``Addition by subtraction ...'' title of the paper series\cite{VRG:bartlett:2006:JCP,VRG:musial:2007:JCP} introducing these methods is remarkable.
\item {\bf Tensor Algebra Interpreter.} Instead of compiling tensor expressions into some form of tensor algebra code (\code{TiledArray}, \code{numpy}) \code{SeQuant} instead supports direct interpretation of its representation of tensor expressions. This allows direct evaluation of the tensor expressions  using \code{TiledArray} as the implementation backend, thereby enabling tensor storage and computation to take advantage of distributed-memory platforms.
The interpreter evaluates tensor networks (i.e., individual terms/diagrams in CC equations) by sequences of binary tensor contractions; the optimal order of contractions is determined to minimize the FLOP count, thus guaranteeing the correct asymptotic scaling (e.g., $\bigo{N^8}$ for 3CC and CCSDT). The interpreter also performs global analysis of all tensor networks involved in the given set of CC amplitude equations to determine common subexpressions that appear among them. Further details of optimization relies of heuristics whose details will be described in the forthcoming \code{SeQuant} manuscript.
\item {\bf MO Integral DSL.} The \code{SeQuant} interpreter invokes user-provided code that maps tensors in the expression to its concrete numerical representation as a \code{TiledArray}'s distributed array object. To simplify implementation of the potentially many types of MO integrals that one encounters in modern many-body electronic structure methods, especially those involving explicit correlation\cite{VRG:klopper:2006:IRPC,VRG:kong:2012:CR,VRG:ten-no:2012:WIRCMS,VRG:hattig:2012:CR} and local correlation,\cite{VRG:werner:2003:JCP,VRG:neese:2009:JCP,VRG:pinski:2015:JCP} \code{MPQC} provides a pseudo-language for describing AO/MO integrals, optionally involving density fitting (DF). This language is used to construct automatically the integral tensors encountered in the tensor expressions generated by \code{SeQuant}. This greatly simplifies interpretation of \code{SeQuant}-generated tensor expression; for example, this allows automated refactorization of equations when using density fitting for approximating MO integrals as well as lazy on-the-fly DF reconstruction of MO integrals that significantly accelerates CCSD on distributed-memory platforms as we described earlier.\cite{VRG:peng:2019:IJQC}
\end{itemize}

Implementation of the closed- and open-shell arbitrary standard CC models as well as $n$CC models (up to 4CC) has been developed in the \code{CCk} class of the developmental version of \code{MPQC}. The essential point about our development toolchain is that all symbolic transformations are performed {\em online}, i.e. during the execution of the user-specified computation task, not offline during code generation. In combination with the {\em online} interpretation of the tensor expressions this allows implementation of arbitrary-order CC methods, similar to the pioneering work of Kallay\cite{VRG:kallay:2001:JCP} and Hirata\cite{VRG:hirata:2000:CPL} but using ordinary dense tensors for storage and computation. Few remaining technical limitations prevent efficient applications beyond quadruples, which is for our current purposes is satisfactory.

\section{Results}\label{sec:results}
To test our implementation of general and approximate coupled cluster methods, we evaluated the total energies of all species (except H and H$_2$) in the High accuracy Extrapolated {\em Ab initio} Thermochemistry (HEAT) benchmark.\cite{VRG:tajti:2004:JCP} 
The HEAT protocols (original and its refinements\cite{VRG:bomble:2006:JCP,VRG:harding:2008:JCP}) have no empirical scaling factors or experimental parameters except for the extrapolation of HF and correlation energies to calculate the complete basis set values.
The HEAT model of the enthalpy at 0K is given by:\cite{VRG:tajti:2004:JCP}
\begin{equation}
\label{HEAT_formula}
\begin{split}
    E_{\text{HEAT}} \equiv & E_{\text{HF}}^\mathrm{CBS} + \delta  E_{\text{CCSD(T)}}^\mathrm{CBS} + \delta E_{\text{CCSDT}}^\mathrm{CBS} + \delta E_{\text{CCSDTQ}} \\
    + & \delta E_{\text{ZPVE}} + \delta E_{\text{REL}} + \delta E_{\text{SO}} + \delta E_{\text{DBOC}},
\end{split}
\end{equation}
where the sum of first four terms (ordered in decreasing magnitude) approximate the exact nonrelativistic electronic energy and the last four terms (in the order of decreasing magnitude) account for the contributions from zero-point vibrational energy, scalar and spin-orbit relativistic effects, and adiabatic effects (diagonal Born-Oppenheimer correction), respectively. 
In the standard HEAT protocol closed- and open-shell species use spin-restricted (RHF) and spin-unrestricted (UHF) reference wave functions, respectively.

This article is focused on accurate estimation of the post-CCSD(T) correlation energy contributions. These are accounted by the 3rd and 4th terms on the right-hand side of \cref{HEAT_formula}. The former accounts for the difference between the perturbative treatment of triples in CCSD(T) and their self-consistent treatment in CCSDT; it is defined as
\begin{equation}
\label{eq:delta-E-CCSDT}
    \delta E_{\text{CCSDT}}^\mathrm{CBS} \equiv E_{\text{CCSDT}}^{\text{TQ}}(\text{fc}) - E_{\text{CCSD(T)}}^{\text{TQ}}(\text{fc}),
\end{equation}
where ``fc'' denotes the frozen-core approximation (only the valence electrons are correlated) and ``TQ'' refers to the correlation energies obtained with the cc-pVTZ and cc-pVQZ basis sets\cite{VRG:dunning:1989:JCP}
and extrapolated to the CBS limit via the inverse cubic formula.\cite{VRG:helgaker:1997:JCP,VRG:halkier:1998:CPL}
The residual electron correlation effects are accounted by the quadruples self-consistently incorporated in the CCSDTQ model:
\begin{equation}
\label{eq:delta-E-CCSDTQ}
    \delta E_{\text{CCSDTQ}} = E_{\text{CCSDTQ}}^{\text{D}}\text{(fc)} - E_{\text{CCSDT}}^{\text{D}}\text{(fc)},
\end{equation}
with ``D'' denoting the correlation energies obtained using the cc-pVDZ basis.

In this work we considered whether the 3CC model can be used to economically approximate the total of high-order corrections in high-end thermochemistry protocols like HEAT. Thus first we examined how accurately the 3CC model can approximate absolute electronic energies and atomization energies obtained with reference high-order models (CCSDTQ and CCSDTQP) using small (cc-pVDZ) basis sets compared to other models including triples and quadruples. The comparison leveraged existing energies for the HEAT dataset from Ref. \citenum{AM:Bomble:2006:}. The results are shown in \cref{tab:heat_1,tab:AE_comparison}, respectively.
For the absolute correlation energies (\cref{tab:heat_1}) the mean unsigned errors (MAE) and their per-valence-electron counterparts (MAEe) of the \{CCSD(T), CCSDT, 3CC, CCSDT[Q], and CCSDT(Q)\} models relative to the CCSDTQ reference are \{1.216, 0.750, 0.261, 0.186, 0.114\} $m E_\mathrm{h}$ and \{0.122, 0.070, 0.024, 0.017, 0.011\} $m E_\mathrm{h}$, respectively. 
The corresponding \{CCSD(T), CCSDT, 3CC, CCSDT[Q], CCSDT(Q), CCSDTQ\} errors vs the CCSDTQP reference are \{1.248, 0.775, 0.270, 0.198, 0.055, 0.061\} $m E_\mathrm{h}$ and \{0.128, 0.074, 0.026, 0.020, 0.006, 0.006\} $m E_\mathrm{h}$, respectively.
The MAEs of atomization energies (\cref{tab:AE_comparison}) predicted by the \{CCSD(T), CCSDT, 3CC, CCSDT[Q], CCSDT(Q)\} MAE vs the CCSDTQ reference are \{2.58, 2.00, 0.46, 0.53, 0.37\} kJ mol$^{-1}$. Similarly, the \{CCSD(T), CCSDT, 3CC, CCSDT[Q], CCSDT(Q), CCSDTQ\} MAE relative to the CCSDTQP reference are \{2.73, 2.08, 0.48, 0.56, 0.19, 0.18\} kJ mol$^{-1}$. 
The data suggests several findings:
\begin{itemize}
    \item For absolute correlation energies and atomization energies 3CC is a {\em drastic} improvement on both the cheaper CCSD(T) model and the more rigorous CCSDT model that has the same $\mathcal{O}(N^8)$ asymptotic cost as 3CC. The improvements are systematic, and are observed for both closed- and open-shell systems.
    \item 
    Although the performance of 3CC for absolute energies is worse than that of the more expensive, $\mathcal{O}(N^9)$, CCSDT[Q] and CCSDT(Q) models, its performance for atomization energies
    is close to that of CCSDT[Q] and only a notch below that of CCSDT(Q).
    \item Performance of CCSD(T) relative to CCSDT is reasonably close for closed-shell systems, but for open-shell systems it is generally worse than CCSDT and much worse than 3CC across the board.
\end{itemize}

\begin{table}[ht!]
    \centering
    \resizebox{\columnwidth}{!}{\begin{tabular}{c|rrrrr|rrrrrr}
    \hline \hline
    & \multicolumn{5}{c|}{$E(\mathrm{CCSDTQ})- E(X)$} & \multicolumn{6}{c}{$E(\mathrm{CCSDTQP})- E(X)$} \\
    Species \ $X$ &	(T)	&	T	&	3CC	&	[Q]	&	(Q)	&	(T)	&	T	&	3CC	&	[Q]	&	(Q)	&	Q	\\ \hline
C & -0.340& -0.029& 0.004& -0.009& -0.007 & -0.340& -0.029& 0.004& -0.009& -0.007& 0.000 \\
\ce{C2H2} & -1.305& -0.911& 0.284& -0.224& 0.109 & -1.429& -1.035& 0.160& -0.348& -0.015& -0.124 \\
CCH & -2.543& -0.930& 0.070& -0.319& 0.048 & -2.661& -1.048& -0.048& -0.437& -0.070& -0.118 \\
CF & -1.188& -0.522& -0.454& -0.607& 0.079 & -1.171& -0.505& -0.437& -0.590& 0.096& 0.017 \\
CH & -0.585& -0.075& 0.049& -0.035& -0.014 & -0.587& -0.077& 0.047& -0.037& -0.016& -0.002 \\
\ce{CH2} & -0.452& -0.081& -0.017& -0.031& -0.009 & -0.455& -0.084& -0.020& -0.034& -0.012& -0.003 \\
\ce{CH3} & -0.521& -0.119& 0.018& -0.046& -0.007 & -0.527& -0.125& 0.012& -0.052& -0.013& -0.006 \\
CN & -3.882& -1.405& 0.024& -0.278& 0.582 & -4.065& -1.588& -0.159& -0.461& 0.399& -0.183 \\
CO & -1.408& -0.951& -0.359& -0.617& 0.140 & -1.461& -1.004& -0.412& -0.670& 0.087& -0.053 \\
\ce{CO2}$^c$ & -1.997 & -1.754 & -1.185 & -1.329 & 0.318 & -- & -- & -- & -- & -- & -- \\
F & -0.191& -0.116& -0.217& -0.004& -0.005 & -0.198& -0.123& -0.224& -0.011& -0.012& -0.007 \\
\ce{F2} & -1.806& -1.536& -0.902& 0.062& 0.158 & -1.881& -1.611& -0.977& -0.013& 0.083& -0.075 \\
\ce{H2O} & -0.617& -0.455& -0.356& -0.019& 0.028 & -0.632& -0.470& -0.371& -0.034& 0.013& -0.015 \\
\ce{H2O2} & -1.552& -1.280& -0.486& -0.042& 0.134 & -1.635& -1.363& -0.569& -0.125& 0.051& -0.083 \\
HCN & -1.431& -1.230& 0.019& -0.152& 0.202 & -1.589& -1.388& -0.139& -0.310& 0.044& -0.158 \\
HCO & -1.617& -0.930& -0.203& -0.493& 0.182 & -1.678& -0.991& -0.264& -0.554& 0.121& -0.061 \\
HF & -0.481& -0.392& -0.512& -0.013& 0.020 & -0.491& -0.402& -0.522& -0.023& 0.010& -0.010 \\
HNO & -1.783& -1.470& -0.192& -0.166& 0.194 & -1.917& -1.604& -0.326& -0.300& 0.060& -0.134 \\
\ce{HO2} & -1.815& -1.110& -0.268& -0.165& 0.156 & -1.899& -1.194& -0.352& -0.249& 0.072& -0.084 \\
N & -0.171& -0.041& -0.037& -0.010& -0.009 & -0.171& -0.041& -0.037& -0.010& -0.009& 0.000 \\
\ce{N2} & -1.532& -1.457& -0.168& -0.058& 0.228 & -1.713& -1.638& -0.349& -0.239& 0.047& -0.181 \\
NH & -0.407& -0.125& -0.075& -0.029& -0.016 & -0.410& -0.128& -0.078& -0.032& -0.019& -0.003 \\
\ce{NH2} & -0.559& -0.223& -0.077& -0.040& -0.006 & -0.569& -0.233& -0.087& -0.050& -0.016& -0.010 \\
\ce{NH3} & -0.584& -0.315& -0.048& -0.041& 0.015 & -0.603& -0.334& -0.067& -0.060& -0.004& -0.019 \\
NO & -1.827& -1.255& -0.204& -0.251& 0.214 & -1.959& -1.387& -0.336& -0.383& 0.082& -0.132 \\
O & -0.190& -0.078& -0.105& -0.010& -0.009 & -0.193& -0.081& -0.108& -0.013& -0.012& -0.003 \\
\ce{O2} & -1.921& -1.730& -0.567& 0.043& 0.178 & -2.089& -1.898& -0.735& -0.125& 0.010& -0.168 \\
OF & -2.104& -0.985& -0.399& -0.282& 0.220 & -2.166& -1.047& -0.461& -0.344& 0.158& -0.062 \\
OH & -0.450& -0.259& -0.258& -0.031& -0.010& -0.456& -0.265& -0.264& -0.037& -0.016& -0.006\\ \hline
MAE & 1.216 & 0.750 & 0.261 & 0.186 & 0.114 & 1.248 & 0.775 & 0.270 & 0.198 & 0.055 & 0.061 \\
MaxAE & 3.882 & 1.754 & 1.185 & 1.329 & 0.582 & 4.065& 1.898& 0.977& 0.670& 0.399& 0.183 \\
MAEe$^{a}$ & 0.122 & 0.070 & 0.024 & 0.017 & 0.011 & 0.128& 0.074& 0.026& 0.020& 0.006& 0.006 \\
MaxAEe$^{b}$ & 0.431 & 0.156 & 0.074 & 0.083 & 0.065 & 0.452& 0.176& 0.070& 0.067& 0.044& 0.020 \\
\hline \hline   
\end{tabular}}
    $^{a}$ Mean absolute error per valence electron ($mE_\mathrm{h}/$e). \\
    $^{b}$ Maximum absolute error per valence electron ($mE_\mathrm{h}/$e). \\
    $^{c}$ The cc-pVDZ CCSDTQP energy not provided in Ref.  \citenum{AM:Bomble:2006:}.
    \caption{Errors in cc-pVDZ valence correlation energies ($mE_\mathrm{h}$) relative to the CCSDTQ and CCSDTQP reference value.}
    \label{tab:heat_1}
\end{table}

\begin{table}[ht!]
\begin{tabular}{c|rrrrr|rrrrrr}
    \hline \hline
    & \multicolumn{5}{c|}{$E(\mathrm{CCSDTQ})- E(X)$} & \multicolumn{6}{c}{$E(\mathrm{CCSDTQP})- E(X)$} \\
    Species &	(T)	&	T	&	3CC	&	[Q]	&	(Q)	&	(T)	&	T	&	3CC	&	[Q]	&	(Q)	&	Q	\\ \hline
\ce{C2H2} & 1.64& 2.24& -0.72& 0.54& -0.32 & 1.97& 2.57& -0.40& 0.87& 0.00& 0.33 \\
CCH & 4.89& 2.29& -0.16& 0.79& -0.16 & 5.20& 2.60& 0.15& 1.10& 0.15& 0.31 \\
CF & 1.72& 0.99& 0.63& 1.56& -0.24 & 1.66& 0.93& 0.57& 1.50& -0.30& -0.06 \\
CH & 0.64& 0.12& -0.12& 0.07& 0.02 & 0.65& 0.13& -0.11& 0.07& 0.02& 0.01 \\
\ce{CH2} & 0.29& 0.14& 0.06& 0.06& 0.01 & 0.30& 0.14& 0.06& 0.07& 0.01& 0.01 \\
\ce{CH3} & 0.48& 0.24& -0.04& 0.10& 0.00 & 0.49& 0.25& -0.02& 0.11& 0.02& 0.02 \\
CN & 8.85& 3.51& -0.15& 0.68& -1.57 & 9.33& 3.99& 0.33& 1.16& -1.09& 0.48 \\
CO & 2.31& 2.22& 0.68& 1.57& -0.41 & 2.44& 2.35& 0.81& 1.70& -0.28& 0.13 \\
\ce{CO2} & 3.35 & 4.12 & 2.57 & 3.41 & -0.90 & -- & -- & -- & -- & -- & -- \\
\ce{F2} & 3.74& 3.42& 1.23& -0.18& -0.44 & 3.90& 3.58& 1.39& -0.02& -0.28& 0.16 \\
\ce{H2O} & 1.12& 0.99& 0.66& 0.02& -0.10 & 1.15& 1.02& 0.69& 0.06& -0.07& 0.03 \\
\ce{H2O2} & 3.08& 2.95& 0.72& 0.06& -0.40 & 3.28& 3.15& 0.93& 0.26& -0.20& 0.20 \\
HCN & 2.42& 3.05& -0.14& 0.35& -0.57 & 2.83& 3.46& 0.28& 0.76& -0.16& 0.41 \\
HCO & 2.85& 2.16& 0.27& 1.24& -0.52 & 3.01& 2.31& 0.42& 1.40& -0.37& 0.15 \\
HF & 0.76& 0.72& 0.77& 0.02& -0.07 & 0.77& 0.73& 0.78& 0.03& -0.06& 0.01 \\
HNO & 3.73& 3.55& 0.13& 0.38& -0.56 & 4.08& 3.89& 0.48& 0.73& -0.21& 0.34 \\
\ce{HO2} & 3.77& 2.50& 0.15& 0.38& -0.46 & 3.97& 2.71& 0.36& 0.59& -0.25& 0.20 \\
\ce{N2} & 3.12& 3.61& 0.25& 0.10& -0.65 & 3.60& 4.09& 0.72& 0.57& -0.17& 0.48 \\
NH & 0.62& 0.22& 0.10& 0.05& 0.02 & 0.63& 0.23& 0.11& 0.06& 0.03& 0.01 \\
\ce{NH2} & 1.02& 0.48& 0.11& 0.08& -0.01 & 1.04& 0.50& 0.13& 0.11& 0.02& 0.03 \\
\ce{NH3} & 1.08& 0.72& 0.03& 0.08& -0.06 & 1.13& 0.77& 0.08& 0.13& -0.01& 0.05 \\
NO & 3.85& 2.98& 0.16& 0.61& -0.61 & 4.19& 3.32& 0.50& 0.95& -0.27& 0.34 \\
\ce{O2} & 4.05& 4.13& 0.94& -0.17& -0.51 & 4.47& 4.56& 1.36& 0.26& -0.09& 0.43 \\
OF & 4.52& 2.08& 0.20& 0.70& -0.61 & 4.66& 2.21& 0.34& 0.84& -0.48& 0.14 \\
OH & 0.68& 0.48& 0.40& 0.06& 0.00& 0.69& 0.48& 0.41& 0.06& 0.01& 0.01\\\hline
MAE & 2.58 & 2.00 & 0.46 & 0.53 & 0.37 & 2.73& 2.08& 0.48& 0.56& 0.19& 0.18 \\
MaxAE & 8.85& 4.13& 2.57 & 3.41 & 1.57& 9.33& 4.56& 1.39& 1.70& 1.09& 0.48\\ 
\hline\hline
\end{tabular}
\caption{Errors in cc-pVDZ atomization enthalpies (1 kJ/mol = 1/2625.4976 $E_\mathrm{h}$) relative to the CCSDTQ and CCSDTQP reference values.}
\label{tab:AE_comparison}
\end{table}

Clearly, our data suggests that high-end ground-state thermochemistry models that go beyond CCSDT should use 3CC in its place.
A more practically relevant question, however, is whether lower-end thermochemistry protocols that go beyond CBS CCSD(T) should aim for the 3CC model as the highest-end treatment of correlation.
To this end we examined how accurately the CBS 3CC model can approximate the exact nonrelativistic energy compared to CBS CCSD(T) and CBS CCSDT. The HEAT electronic energy (the first 4 contributions in \cref{HEAT_formula}) was again used as the reference. Since the latter includes the CBS CCSD(T) and CBS CCSDT energies as partial summands, the errors
of CBS CCSD(T) and CBS CCSDT are obtained from the HEAT components (\cref{eq:delta-E-CCSDT,eq:delta-E-CCSDTQ}):
\begin{align}
\label{eq:Delta-E-CCSD(T)-CBS}
    \Delta E_\mathrm{CCSD(T)}^\mathrm{CBS} \equiv & \delta E_\mathrm{CCSDT}^\mathrm{CBS} + \delta E_\mathrm{CCSDTQ}\\
\label{eq:Delta-E-CCSDT-CBS}
    \Delta E_\mathrm{CCSDT}^\mathrm{CBS} \equiv & \delta E_\mathrm{CCSDTQ}.
\end{align}
By estimating the CBS limit of 3CC using the same TQ extrapolation protocol used for CCSDT we obtain the error of CBS 3CC relative to the HEAT energy as follows:
\begin{align}
\label{eq:Delta-E-3CC-CBS}
  \Delta E_\mathrm{3CC}^\mathrm{CBS} \equiv & \delta E_\mathrm{CCSDTQ} + (E_\mathrm{3CC}^\mathrm{TQ}(\mathrm{fc}) - E_\mathrm{CCSDT}^\mathrm{TQ}(\mathrm{fc}))
\end{align}

To simplify the analysis we focused on the atomization energies only (see \cref{tab:AE_CBS_errors}).
The mean signed and absolute errors for \{CCSD(T), CCSDT, 3CC\} are \{0.96, 2.00, -0.27\}
 and \{1.07, 2.00, 0.52\} kJ/mol, respectively. 
The maximum absolute errors (MaxAE) for \{CCSD(T), CCSDT, 3CC\} are \{6.64, 4.13, 1.44\} kJ/mol. 
Clearly, 3CC at the CBS limit predicts substantially more accurate atomization energies than either CCSD(T) or CCSDT. Thus it is the recommended post-CCSD(T) heuristic for ground-state thermochemistry. Since 3CC provided most accurate triples-only absolute correlation energies and atomization energies when using with a small (double-zeta) basis, 3CC should be considered to be the recommended iterative triples heuristic for other uses, such as the starting point for incorporation of quadruples (e.g., in the context of the $\mathcal{O}(N^9)$ CCSDT(Q) model).

\begin{table}[ht!]
\begin{tabular}{crrr}
\hline\hline
 & CCSD(T) & CCSDT & 3CC \\ \hline
MSE     &   0.96   &   2.00   &  -0.27 \\
MAE	    &   1.07    &   2.00    &   0.52 \\
MaxAE   &	6.64	&   4.13	&   1.44 \\
\hline \hline
\end{tabular}
\caption{Statistical averages of the errors in CBS CCSD(T) [\cref{eq:Delta-E-CCSD(T)-CBS}], CCSDT [\cref{eq:Delta-E-CCSDT-CBS}], and 3CC [\cref{eq:Delta-E-3CC-CBS}] atomization energies (kJ/mol) relative to the HEAT electronic energy reference.}
\label{tab:AE_CBS_errors}
\end{table}

Comparison of the CBS data in \cref{tab:AE_CBS_errors} vs its cc-pVDZ counterpart in \cref{tab:AE_comparison} reveals another puzzle. Whereas in a small basis CCSD(T) was less accurate than CCSDT, at the CBS limit CCSD(T) is more accurate {\em on average} than CCSDT (although the maximum error of the former is larger). This leads to several conclusions. The basis set error of the difference of CCSDT and CCSD(T) atomization energies is nonnegligible (1-2 kJ/mol) with the cc-pVDZ basis.
The corresponding basis set error of the 3CC-CCSDT energies is substantially smaller on average, but still not entirely negligible. 
This illustrates that the assessment of various triples-including models cannot utilize the minimally-viable cc-pVDZ basis. Of course, we also need a more thorough assessment of the basis set convergence of the effects of quadruples and higher-body clusters to be able to assess even the triples-including models at the CBS limit.

\section{Summary and Perspective}
\label{sec:summary}

The 3CC method is an internally corrected approximation to the CCSDT.
We presented an automated implementation of this method for both closed- and open-shell species and applied it to the well-known HEAT benchmark dataset.
In agreement with our previous findings on a small set of closed-shell systems,\cite{VRG:rishi:2019:JCP} the 3CC model with $\mathcal{O}(N^8)$ complexity predicts far more accurate (in reference to the CCSDTQ and CCSDTQP models) valence correlation energies and atomization energies than both the $\mathcal{O}(N^7)$ complexity CCSD(T) heuristic as well as the formally more rigorous CCSDT model that has the same $\mathcal{O}(N^8)$ complexity. For atomization energies these observations hold both with smallest viable (cc-pVDZ) basis as well as in the complete basis set limit.
For example, the mean and maximum absolute errors of CBS CCSD(T), CCSDT, and 3CC atomization energies relative to the HEAT electronic contribution reference are \{1.07, 2.00, 0.52\} and \{6.64, 4.13, 1.44\} kJ/mol, respectively. This provides strong hints that the 3CC model is not only the recommended post-CCSD(T) heuristic for high-end studies of chemical energitics, but should be considered as the recommended iterative triples heuristic more broadly, such as the starting point for incorporation of quadruples (e.g., in the context of the $\mathcal{O}(N^9)$ CCSDT(Q) model).

Although application of high-end model like 3CC, with its $\mathcal{O}(N^8)$ cost complexity, seems hardly practical for more than a few atoms, there are many viable ways to reduce its cost. For large basis sets the particle-particle ladder dominates the formal cost of coupled-cluster methods, and it is by now known how to reduce the formal complexity and overall cost of ladder-type diagrams even for relatively small systems by factorization of the Coulomb integral tensor using pseudospectral techniques,\cite{VRG:friesner:1985:CPL} tensor hypercontraction (THC),\cite{VRG:hohenstein:2012:JCP} or better yet using robust canonical polyadic decomposition combined with density fitting.\cite{VRG:pierce:2021:JCTC} Factorization of the cluster amplitudes themselves can lead to further complexity reductions and lead to realizable savings as shown recently for CCSDT model.\cite{VRG:lesiuk:2020:JCTC}
Lastly, iterative evaluation of the (T) correction in the context of PNO-based CC methods suggests that amplitude sparsification and compression can make CC methods with iterative treatment of triples, such as 3CC, practical.

\begin{acknowledgement}

We thank Dr. Varun Rishi and Dr. Ashutosh Kumar for useful discussions.

This work was supported by the U.S.~Department of Energy (DOE), Office of Science, Office of Advanced Scientific Computing Research and Office of Basic Energy Sciences, Scientific Discovery through the Advanced Computing (SciDAC) program under award DE-SC0022263. The development of the \code{SeQuant} and \code{TiledArray} software packages is supported by the U.S. National Science Foundation, Office of Advanced Cyberinfrastructure under award 2103738. The Advanced Research Computing (ARC) at Virginia Tech provided the computational resources and support that contributed to this work.

\end{acknowledgement}

\bibliography{vrgrefs, nakulrefs, ajayrefs}

\end{document}